\documentclass{ws-procs9x6A}

\newcommand{\be}{\begin{equation}} \newcommand{\ee}{\end{equation}}
\newcommand{\ba}{\begin{eqnarray}} \newcommand{\ea}{\end{eqnarray}}
\newcommand{\bea}{\begin{eqnarray}} \newcommand{\eea}{\end{eqnarray}}
\newcommand{\bean}{\begin{eqnarray*}} \newcommand{\eean}{\end{eqnarray*}}
\newcommand{\s}[1]{{\scriptscriptstyle #1}}
\newcommand{\st}{{\scriptscriptstyle T}}
\def\slash{\rlap{/}}

\begin{document}

%\title{Universality of Single Spin Asymmetries in Hard Processes}
\title{Universality of Single Spin Asymmetries in Hard Processes\footnote{
Contributed paper at DIS2006, April 20-24, 2006, Tsukuba, Japan}}

\author{P.~J. MULDERS, C.~J. BOMHOF and F. PIJLMAN} 

\address{Department of Theoretical Physics, Faculty of Sciences,\\
De Boelelaan 1081, NL-1081 HV Amsterdam, the Netherlands\\ 
E-mail: mulders@few.vu.nl}

\maketitle

\abstracts{
We discuss the use of time reversal symmetry in the classification
of parton correlators. Specifically, we consider the role
of (small) intrinsic transverse momenta in these correlators
and the determination of the proper color gauge link. 
The transverse momentum weighted correlators in hard processes can
be expressed as a product of universal gluonic pole matrix elements and
gluonic pole cross sections.}

%\subsection*{Introduction}

For (semi)-inclusive measurements, the cross sections in hard
scattering processes factorize into a hard squared amplitude and
distribution and fragmentation functions. These functions enter in forward
matrix elements of nonlocal combinations of quark and gluon field 
operators. Generically we need for the distribution functions the
(light-cone) correlator
\be
\Phi(x) = \left. \int \frac{d(\xi\cdot P)}{2\pi}\ e^{i\,p\cdot\xi}
\,\langle P\vert \phi^\dagger(0)\,\phi(\xi)\vert P\rangle\right|_{
\xi\cdot n = \xi_\st = 0},
\label{corr-lc}
\ee
where $\phi$ can be the quark field or the gluon field strength. The
correlator depends on the momentum fraction appearing in the Sudakov 
decomposition of the quark momentum $p = x\,P + p_\st + \sigma\,n$,
where $n$ is an (arbitrary) light-like vector for which $P\cdot n$ is
of the order of the hard scale (say $\sqrt s$). 
Of particular interest in our study is the dependence on transverse 
momenta (with respect to $P$ and $n$), appearing in the 
light-front correlators
\be
\Phi(x,p_\st) = \left. \int \frac{d(\xi\cdot P)\,d^2\xi_\st}{(2\pi)^3}
\ e^{i\,p\cdot\xi}
\,\langle P\vert \phi^\dagger(0)\,\phi(\xi)\vert P\rangle\right|_{
\xi\cdot n = 0}.
\label{corr-lf}
\ee
In a similar way correlators relevant for fragmentation functions can
be written down. The actual distribution (and fragmentation) functions
appear in the parametrization of the correlators and will be discussed
below.

%\subsection*{Single spin asymmetries and time reversal invariance}

QCD is invariant under time reversal (T). This means that observables
can be characterized by their T-behavior. Single spin asymmetries (SSA), 
i.e.\ differences of cross sections in which only one of the (initial
or final state) spins is flipped, are examples of T-odd observables.
Also the correlators, in particular those for distribution functions
mentioned in the previous section, can be divided into T-even and T-odd
parts. The T-behavior of the states $\vert P\rangle$ and that of the
quark and gluon field operators is known. The functions appearing in
these respective parts are referred to as T-even and T-odd distribution
functions. For distributions, there are no collinear T-odd functions in 
Eq.~\ref{corr-lc} but there are T-odd transverse momentum dependent
(TMD) distribution functions in the correlator in Eq.~\ref{corr-lf}. 
Since the hard process, at least at tree level, is T-even, one must have
in the description of SSA at least one T-odd function
(or in general an odd number of them).

%\subsection*{Measuring intrinsic transverse momenta}

The collinear part $x$ of the quark momenta with
respect to the hadron momentum appearing in the Sudakov decomposition may 
be obtained from the hard kinematics, e.g. in deep inelastic scattering
$x = x_{\s B} = -q^2/2p\cdot q$. This is also possible for the transverse
momentum by measuring the non-collinearity in the process, e.g. the
transverse momentum of a produced hadron in leptoproduction with
respect to incoming hadron $P$ and the momentum transfer $q$, or the
deviation from back-to-backness of jets in the transverse plane in
hadron-hadron scattering\cite{BV,BBMP}. The correlators that can 
be obtained in a
suitably weighted cross sections are the transverse moments
$\Phi_\partial^\alpha(x) \equiv
\int d^2p_\st\ p_\st^\alpha \Phi(x,p_\st)$,
which again are lightcone dominated.

%\subsection*{An example}

As an example consider the correlator for quarks. In that case the
nonlocal field combination shown in our definitions is given by
$\phi^\dagger(0)\,\phi(\xi) \ \longrightarrow
\ \overline \psi(0)\,{U}_{[0,\xi]}\,\psi(\xi)$,
where the final correlator appearing in a calculation contains
a gauge link ${U}_{[0,\xi]} 
= {P}\,exp\left(-ig\int_0^\xi ds^\mu\,A_\mu\right)$, which
ensures color gauge invariance. The integration path in the 
gauge link follows from a
diagrammatic calculation that includes for a given hadron besides
quark correlators also correlators with in addition
collinear $A\cdot n$ gluons. Such correlators turn out to be
leading as well and the resummed result nicely produces the color 
gauge invariant result. The path also follows from the calculation
to a straight-line path along $n$, indicated as $U^n_{[0,\xi]}$. 

For the TMD correlators, however, the nonlocal operator combinations
involve transverse separation of fields and also the gauge link 
acquires a transverse piece. In the case of electroweak processes
these turned out to be gauge links consisting of two pieces along 
the lightcone connected at light-cone infinity. However, the links 
for an incoming hadron in one-particle-inclusive leptoproduction 
($\ell + H \rightarrow \ell + h$) and
for an incoming hadron in the Drell-Yan process
($H + H^\prime \rightarrow \ell + \bar \ell$) turned out to be
different connected at lightcone $\pm \infty$ (future and past 
pointing gauge links), respectively.
In general, e.g. in hadron-hadron scattering, more complicated, 
but calculable, gauge links appear\cite{BMP04}. 

%\subsection*{Gluonic pole matrix element}

The structure of the gauge links is not relevant if one only deals
with collinear (lightcone) correlators. In the transverse
moments, showing up in azimuthal asymmetries,
it is relevant. They are written as
\be
\Phi_\partial^{[U]\alpha}(x) =
\Phi_\partial^{\alpha}(x) + C_G^{[U]}\,\pi\,\Phi_G^\alpha(x,x),
\label{decom}
\ee
containing a link-independent part $\Phi_\partial^\alpha$, 
involving among others standard twist three collinear functions and a 
part that is the product of a link-dependent color factor $C_G^{[U]}$
and a link-independent  gluonic pole matrix element\cite{BMT,BMP03} 
derived from the quark-gluon correlator
\bea
\pi\,\Phi_G^\alpha(x,x-x^\prime) & = & 
\int \frac{d(\xi\cdot P)}{2\pi}\,\frac{d(\eta\cdot P)}{2\pi}
\ e^{i(x-x^\prime)P\cdot\xi}\,e^{ix^\prime P\cdot \eta}
\nonumber \\ && \mbox{}\qquad
\times\left.\langle P\vert \overline \psi(0)\,U^n_{[0,\eta]}
\,gF^{n\alpha}(\eta)
\,U^n_{[\eta,\xi]}\,\psi(\xi)\vert P\rangle
\right|_{LC},
\label{gluonicpole}
\eea
where LC indicates the restriction to the lightcone. The color factors
directly follow from the link structure. For instance
for the future and past pointing links in leptoproduction and
Drell-Yan one has $C_G^{[\pm]} = \pm 1$.

The time reversal properties of the gluonic pole part in Eq.~\ref{gluonicpole}
is opposite to that of the link-independent contribution or the 
integrated correlator. For instance the TMD quark distributions in
unpolarized hadrons are contained in a quark correlator (including link $U$)
\be
\Phi^{[U]}(x,p_\st) = \frac{1}{2}\left\{
f_1(x,p_\st^2)\slash P 
+ h_1^\perp(x,p_\st^2)\,\frac{i[\slash p_\st,\slash P]}{2M}\right\}.
\ee
In this expression the functions would depend on the link. If we work
with the integrated correlator and the transverse moments, we can
put the link dependence in the coefficients $C_G^{[U]}$ and universal
matrix elements, the integrated ones
$\Phi(x) = \tfrac{1}{2}\,f_1(x)\,\slash P$,
the link-independent part of the transverse moment, which is zero,
for unpolarized hadrons, $\Phi_\partial^\alpha(x) = 0$,
and the gluonic pole matrix element
$\pi\,\Phi_G^\alpha(x,x) 
= \tfrac{1}{4}\,ih_1^{\perp (1)}(x)\,[\slash P,\gamma^\alpha]$,
where $h_1^{\perp (1)}(x)$ is the $\bm p_\st^2/2M^2$-weighted function.
Thus there is a universal function $h_1^{\perp (1)}$ which, because of
the link-dependent factors $C_G^{[\pm]}$ in Eq.~\ref{decom}, 
appears with opposite signs in leptoproduction and Drell-Yan scattering.
 
%\subsection*{Gluonic pole cross sections}

Beyond the simple electroweak processes, one finds in general that 
several diagrams contribute to the hard scattering part. For instance
for the quark-quark scattering contributing to pp-scattering one has 
for identical quarks at lowest order already four contributions
\be
\hat \sigma_{qq\rightarrow qq} 
= \sum_{[D]} \hat \sigma^{[D]}_{qq\rightarrow qq}.
\ee
It is this cross section that is multiplied by distribution and fragmentation
functions, e.g. the function $f_1^q(x)$ in unpolarized scattering.
For azimuthal asymmetries, one finds that each
contribution in principle leads to a particular link in the correlator
that connects partons in the hard part to hadrons and correspondingly each
contribution has its specific strength $C_G^{[D]}$. The links and factors
depend on the color flow through the diagram.
In order to accomodate these, it is convenient to define gluonic pole
cross sections
\be
\hat \sigma_{[q]q\rightarrow qq} 
= \sum_{[D]} C_G^{[D]}\,\hat \sigma^{[D]}_{qq\rightarrow qq}.
\ee
These cross sections will appear multiplied with a distribution
function in the transverse moments, which for a particular 
T-odd SSA might be the function $h_1^{\perp (1) q}(x)$.

A treatment\cite{BBMP} of the quark contributions in pp-scattring
along these lines has been completed, while also the general procedure 
to find gauge links in hard processes has been given\cite{BMP06}. 
In addition, the transition from small to large transverse momenta in
such processes requires care\cite{EKT}.

\section*{Acknowledgments}
This work is included in the research programme of the EU Integrated 
Infrastructure Initiative Hadron Physics (RII3-CT-2004-506078). 
The work of CJB is supported by the Foundation for Fundamental 
Research of Matter (FOM) and the National Organization for Scientific 
Research (NWO).

\end{document}